\documentclass[12pt,preprint]{emulateapj}

\newcommand{\kms}{km\ s$^{-1}$}

\newcommand{\Msun}{$M_{\sun}$}

\newcommand{\bm}{beam$^{-1}$}

\slugcomment{Submitted to ApJ}

\shorttitle{GMCs in M31,M33 \& Milky Way}
\shortauthors{Sheth et al.}

\begin{document}
\title{Comparative Analysis of Molecular Clouds in M31, M33 and the Milky Way}

\author{Kartik Sheth\altaffilmark{1,2}}
\affil{Spitzer Science Center, California Institute of Technology, MS 105-24, Pasadena, CA 91125}
\email{kartik@astro.caltech.edu}
\author{Stuart N. Vogel\altaffilmark{2}}
\affil{Department of Astronomy, University of Maryland, College Park, MD 20742-2421}

\email{vogel@astro.umd.edu}

\author{Christine D. Wilson}

\affil{Department of Physics and Astronomy, McMaster University,
  Hamilton, ON, Canada, L8S 4M1}

\email{wilson@physics.mcmaster.ca}
\and
\author{T. M. Dame}
\affil{Harvard-Smithsonian CfA,Mail Stop 72, 60 Garden Street, Cambridge, MA 02138 }
\email{tdame@cfa.harvard.edu}

\begin{abstract}
  
  We present BIMA observations of a 2$\arcmin$ field in the
  northeastern spiral arm of M31.  In this region we find six giant
  molecular clouds that have a mean diameter of 57$\pm$13 pc, a mean
  velocity width of 6.5$\pm$1.2 \kms, and a mean molecular mass of 3.0
  $\pm$ 1.6 $\times$ 10$^5$\Msun.  The peak brightness temperature of
  these clouds ranges from 1.6--4.2 K.  We compare these clouds to
  clouds in M33 observed by \citet{wilson90} using the OVRO 
  millimeter array, and some cloud complexes in the Milky Way
  observed by \cite{dame01} using the CfA 1.2m telescope.  In order to
  properly compare the single dish data to the spatially filtered
  interferometric data, we project several well-known Milky Way
  complexes to the distance of Andromeda and simulate their
  observation with the BIMA interferometer.  We compare the simulated
  Milky Way clouds with the M31 and M33 data using the same cloud
  identification and analysis technique and find no significant
  differences in the cloud properties in all three galaxies.  Thus we
  conclude that previous claims of differences in the molecular cloud
  properties between these galaxies may have been due to differences
  in the choice of cloud identification techniques.  With the upcoming
  CARMA array, individual molecular clouds may be studied in a variety
  of nearby galaxies.  With ALMA, comprehensive GMC studies will be
  feasible at least as far as the Virgo cluster.  With these data,
  comparative studies of molecular clouds across galactic disks of all
  types and between different galaxy disks will be possible. Our
  results emphasize that interferometric observations combined with
  the use of a consistent cloud identification and analysis technique
  will be essential for such forthcoming studies that will compare GMCs in
  the Local Group galaxies to galaxies in the Virgo cluster.
\end{abstract}

\section{Introduction}
In the inner disk of the Milky Way, the dominant component of the
interstellar medium is molecular gas which is distributed in discrete
cloud complexes \citep{scoville87, scoville90}.  Nearly all star
formation in our Galaxy is associated with these clouds
\citep{scoville87, blitz93}.  Therefore, understanding the
distribution and properties of molecular clouds is a prerequisite for
understanding galaxy evolution.  Studies of the molecular gas emission
in the Milky Way show that while the cloud mass spectrum extends over
several orders of magnitude, a majority of the molecular gas mass is
contained in large, massive cloud complexes, which are typically
$\sim$40 pc in diameter and have masses on the order of 1 $\times$
10$^5$ \Msun; these complexes are usually referred to as giant molecular clouds
(GMCs)\footnote{In this paper, we will use the words giant molecular
clouds, GMCs and cloud complexes interchangeably.}  and are considered
to be the basic organizational unit of the molecular interstellar
medium \citep{scoville90, combes91, young91}.

Compared to the Milky Way, the properties of GMCs in external galaxies are
not well known.  The primary reason for this is a lack of high
resolution, high sensitivity data which are necessary for spatially
resolving individual complexes.  Only in the Magellanic Clouds can
single dish telescopes resolve extragalactic GMCs
\citep{israel93,rubio93, fukui01}.  Even in the nearest spirals (e.g.,
M31 or M33), the angular size of a typical 40 pc GMC is
$\sim$12$\arcsec$, approximately one half the resolution of the IRAM
30m single dish telescope at 2.6mm, the wavelength of the CO (J=1--0)
line emission.  For these galaxies, interferometric observations are
necessary to resolve GMCs.

GMCs have been detected using millimeter interferometers in the two
nearest spiral galaxies, M33 \citep{wilson90,
  engargiola03,rosolowsky07b} and M31
\citep{vogel87,wilson93,loinard98,rosolowsky07a}. In M33,
\citet{wilson90} observed 19 different fields and identified over 30
GMCs but only 9 GMCs were suitable for a study of cloud properties.
The number of GMCs detected in most spiral galaxies has been less than
a dozen, presumably due to the small coverage area. This is beginning
to change.  Recently a survey of M33 was undertaken by the BIMA array.
The angular resolution was coarse ($\sim$13$\arcsec$) but adequate to
identify nearly 150 GMCs (\citealt{engargiola03}, see also
\citealt{rosolowsky07b}).  With the upcoming CARMA array,
ground-breaking studies like the M33 survey are likely to become
routine for external galaxies.

It is important to note that previous studies of GMCs in external
galaxies have used different instruments with varying angular and
linear resolutions and varying sensitivity.  Moreover, these studies
have used different cloud-identification methods to define GMCs and
determine their properties.  Yet, in almost all of these studies, the
observations have detected discrete molecular features similar to
Galactic GMCs. Particularly notable is the consistent linewidth-size
relationship in most of these observations.

However, several studies have noted differences between Galactic GMCs
and those found in some irregular and dwarf galaxies.  In the
irregular galaxies NGC 6822 \citep{wilson94a} and the SMC
\citep{rubio93}, the clouds are slightly less massive, and perhaps
smaller than clouds seen in the Milky Way.  In the LMC,
\citet{cohen88} find that for a given size and line width, clouds are
about six times fainter in CO than comparable clouds in the Milky Way.
The same is true of the dwarf NGC 1569 where the CO-to-H$_2$
conversion factor is reported to be six times larger than the Galactic
value \citep{taylor99}.  In IC10, the cloud sizes and masses are
similar to the Milky Way clouds, but the CO-to-H$_2$ conversion factor
may be twice as high as that seen in the Galaxy if the clouds are
assumed to be self-gravitating \citep{wilson91}.  These studies
suggest that there are noticeable differences between the Galactic
GMCs and those found in irregular or dwarf galaxies, that may be
attributed to the different host environments (e.g., metallicity) of
these galaxies.

GMCs in the three nearest spiral galaxies, M31, M33 and the Milky Way
can be observed at sufficiently high resolution ($\sim$20 pc) to
remove ambiguities from coarse resolution.  Are the GMC properties in
these spiral galaxies fundamentally different?  \citet{wilson90} claim
that no GMC larger than 4$\times$10$^5$ \Msun is seen in M33, and
unlike the Milky Way, 50\% of the molecular mass in M33 resides in
small clouds with masses less than 8$\times$10$^4$ \Msun, based on a
comparison of interferometric and single dish flux from the CO(J=1--0)
line.  On the other hand, analysis of the $^{13}$CO/$^{12}$CO line
ratio for a number of M33 clouds suggests that the total mass measured
from the $^{12}$CO (J=1-0) line may be overestimated and the actual
conversion factor from the CO to H$_2$ mass may be lower in the
diffuse gas; in that case, the amount of mass in GMCs would constitute
most of the molecular mass in M33 \citep{wilson94b}.  With coarser
resolution data, \citet{engargiola03} find a steep GMC mass spectrum
in M33 and report a characteristic mass of 7 $\times$ 10$^4$\Msun.
They conclude that molecular clouds in M33 are formed rapidly from the
atomic gas and are short lived entities.

On the other hand, in M31 \citet{vogel87} and \citet{wilson93} note
that the four GMCs they detect are similar to Galactic GMCs.
\citet{loinard99} smooth a section of the Carina arm in the Milky Way
to the resolution of single dish M31 maps and conclude that there are
general similarities between the distribution of molecular gas in M31
and the Milky Way.  Similar results were also found by \citet{heyer00}
who compared the molecular gas in M31 and the Milky Way using single
dish observations; in their study,\citet{heyer00} find a number of
similarities in the molecular gas proeprties between M31 and the Milky
Way such as the amount of molecular gas at radii larger than 8 kpc,
large arm to interarm contrasts, and similar surface brightness, line
widths and spacings of GMCs in spiral arms.  \citet{rosolowsky07a} has
recently completed a large survey of clouds in M31.  He also finds
that the GMCs in M31 are similar to those found in the Milky Way.
These results depend on comparison of properties derived from datasets
which are obtained with a different instrument and analyzed with
different methods of cloud identification.  The Milky Way data, for
example, come from single dish observations with a signal to noise
that is at least 2-4$\times$ higher than in the M33 and M31 datasets.
The datasets also differ in that single-dish telescopes are inherently
different than the aperture synthesis maps: the interferometer will
miss smooth, extended emission whereas the single-dish data may suffer
from beam dilution.  The Milky Way survey also suffers from varying
linear resolution.  Finally, the
%%
%% method for identifying clouds in these different studies is different.
%%
%% SNV this is the 3rd or 4th time you've talked about varying methods,
%% and I want to make it a little clearer that in this case you are 
%% refering to just the milky way dstudies
methods for identifying clouds in the different Milky Way studies
vary.  These studies (e.g., \citealt{dame86,sanders85}) have usually
projected a typical (l,b,v) cube on one of its axes and used a
integrated intensity contour to identify discrete features they call
clouds or complexes.  The sizes were measured from the total area of
the cloud, assuming that the cloud was spherical \citep{dame86}, or by
measuring the chord across the velocity centroid in a
position-velocity diagram \citep{sanders85}.  In M33, \citep{wilson90}
used a two-tiered selection criterion which required that a cloud be
%% SNV modified the following sentence - check it is still correct
%% I inserted ``or more'' and ``at least''
3$\sigma$ or more in 2 adjacent channels and that its summed flux be at least
3$\sigma$ above the noise.

The lack of consistency in these methods and differences in observing
techniques are worrisome. A proper comparative study of GMCs in these
galaxies requires that the differences described above be eliminated.
This is particularly important in view of future telescopes such as CARMA
and ALMA which will have the ability to mosaic large regions of
galaxies like M31 and M33.  The large scale surveys will have the
detail necessary for in-depth comparisons of the molecular ISM in
Local Group and more distant galaxies.  These future studies will
address the fundamental question of how the molecular ISM varies from
galaxy to galaxy.  Knowing this is critical to understanding star
formation and galaxy evolution.

In this paper, we make an attempt to do a fair comparison of the
molecular emission in the three nearest spiral galaxies, i.e., M31,
Milky Way and M33.  We compare the M31 complexes to the simulated
Milky Way complexes and previous surveys, using a cloud-identification
technique in which we identify complexes using a single integrated
intensity contour (\S \ref{newres}), and discuss how previous
conclusions about M33 would change if the same technique of cloud
identification were used (\S \ref{just}).  An important addition to
this analysis would be the addition of single dish data for the M31
and M33 clouds which would allow for an even more comprehensive
comparison.  Of course, one would still need to project the Milky Way
clouds to the distance of Andromeda to compare the clouds at the same
spatial resolution.  Our method is intended to lay the groundwork for
future studies that will compare molecular clouds across galactic
environments and across different galaxy types with telescopes such as
CARMA, SMA and ALMA.

In \S \ref{methods}, we describe two automated cloud identification
algorithms (Gaussclump developed by \citet{stutzki90}, and Clumpfind
developed by \citet{williams94}), and their advantages and
disadvantages.  While we do not advocate using these algorithms to
describe the characteristics of GMC populations in a galaxy, these
algorithms may be used to compare the M31, M33 and the simulated Milky
Way data.  We describe the results of this experiment in \S \ref{auto}
and discuss our overall conclusions in \S \ref{conc}. Preliminary results from
this work have been published in \citet{sheth00}.

\section{Observations}\label{obs}
\subsection{M31 BIMA Observations}

We observed CO(J=1--0) emission in the northeastern spiral arm of M31
with the BIMA (Berkeley-Illinois-Maryland Association)\footnote{The
  BIMA Array was partially funded by grant AST-9981289 from the
  National Science Foundation} array.  Using a 7-pointing hexagonal
grid (one central pointing surrounded by a ring of six pointings, each
separated by 53$\arcsec$ from its two neighbors in the ring and the
center), we mosaicked a region known to have strong single-dish CO
emission (point 4,0 in \citep{ryden86}).  The mosaicked fields were
observed sequentially with an integration time of 1 minute per field
for 35 minutes, at the end of which, the calibrator was observed for 3
minutes.  The total data presented here was collected in five separate
tracks for a total integration time of 18.2 hours, or 2.6 hrs per
field.  The fields were combined using MIRIAD \citep{sault95} with an
algorithm developed by \citet{sault96}.  Note that the final maps
generated from this algorithm have variable gain and sensitivity. The
noise increases monotonically from the map center towards the edge.
The gain is constant over a fairly large region (in this case, the
central $\sim$180$\arcsec$) and then drops sharply to the edge.  For
our analysis, we only considered the region where the gain was higher
than 0.4.  Further details of the observations are given in
Table~\ref{tbl-1}.

\subsection {M33 and Milky Way Data} 

The M33 survey was conducted by \citet{wilson90} with the OVRO (Owens
Valley Radio Observatory)\footnote{The OVRO Array was partially funded
by grant AST-9981546 from the National Science Foundation}
millimeter array.  The data reduction process is described in
\citet{wilson90}.

The Milky Way observations and reductions are from the 1.2m CfA survey
\citep{dame01}.  The different complexes studied in the present work
have been previously published as follows: Gem OB1 \citep{stacy91},
W3CO \citep{digel96}, Cas A \citep{ungerechts00}, and the Carina arm
\citep{grabelsky87,bronfman89}.  Table \ref{tbl-1a} lists the datasets
and instruments used in this study.

\section{Results: M31 Clouds}\label{newres}
\begin{figure}
\plotone{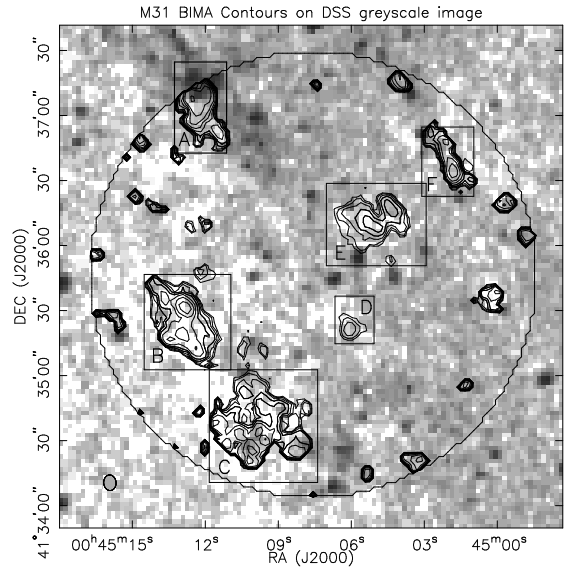}
\caption{Velocity-integrated, primary beam corrected, intensity map of
  the M31 field shown in logarithmic contours of 1.5$^n \times$ 2 K
  km/s, where $n$=-1,1,2,3,...  The grey scale image is a DSS J-band
  image in reverse color, so that white color represents obscuring
  dust.  The six complexes used for comparison to the Milky Way are
  identified with the boxes labelled A-F.  The outer circle is the
  gain=0.4 contour.  \label{fig1}}
\end{figure}

In Figure \ref{fig1}, we show the velocity-integrated CO(J=1--0)
intensity map in contours overlaid on a grey-scale optical\footnote
{Digitized Sky Survey image of M31 obtained from the Skyview database
\citep{mcglynn96}} image.  Our field of view encompasses two dust
lanes
%% SNV SINCE THE DUST LANES AREN'T TOO OBVIOUS, i THOUGHT WE SHOULD 
%% HELP THE READER:
which can be seen trending diagonally across the field from NE to SW.
The majority of the CO emission lies in these two dust lanes, which
are separated by a distance of 2 kpc (assuming an inclination,
i=77$^\circ$).  There are six distinct molecular cloud complexes,
which are labeled A-F.  The complex labeled E lies between the two
dust lanes along a dust spur which runs between the two dust lanes.
We rotated the data cube, collapsed one spatial dimension and
generated RA-velocity and DEC-velocity maps for each complex.  We
checked these maps to verify that each complex is a
well-defined and separate entity.  None of the complexes have emission
separated by more than a few \kms\ except complex C.

%% SNV MODIFIED FOLLOWING SENTENCE CHANGING ``observed'' to ``detected''
%% AND REARRANGED.
\begin{figure}
\plotone{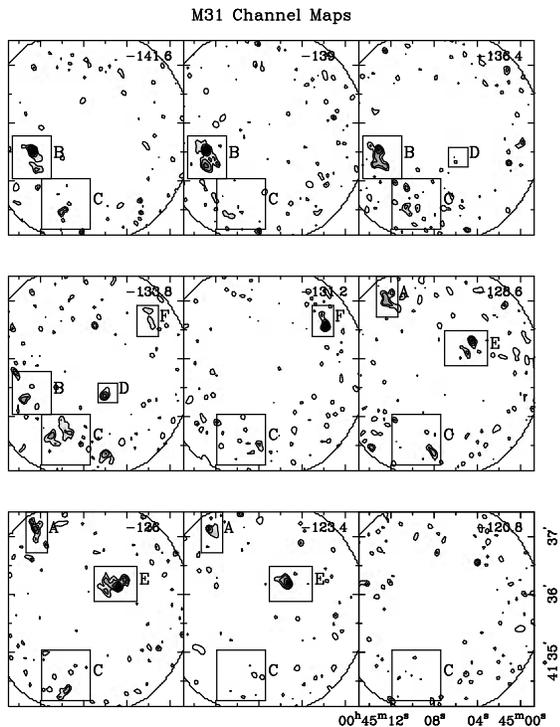}
\caption{Channel maps showing the CO (J=1--0) emission in the
  northeastern spiral arm of M31.  These maps are not primary beam
  corrected.  The contours are logarithmic, 1.5$^n \times$ 0.22 Jy
  beam$^{-1}$.  The boxes show the complexes A-F.  The channel maps
  are 2.6 \kms wide starting at -142 \kms.  The outermost circular
  contour is the gain=0.4 contour. \label{fig2}}
\end{figure}
In Figure \ref{fig2}, we show channel maps in which CO (J=1--0)
emission is detected.  These maps highlight the complex structure
present in these clouds. Complex A, towards the northern edge of our
field of view, is in the dust lane adjacent to a bright HII region.
Further south in the dust lane is complex B which has considerable
structure; it is made up of at least two or three different clumps
which the automated algorithms separate into different clumps.
Complex C is the weakest complex in our maps. It is patchy and extends
over a large region, and over 10 \kms\ in velocity.  The spectrum for
this complex shows that it may be a blend of several clouds.
Therefore we do not calculate the size or velocity width for this
complex. Complex D may be considered a prototypical GMC; it is
well-isolated, i.e. unblended and has nice symmetry.  This complex is
also rather weak but we are fortunate that it is near the center of
our mosaicked field where the noise is the lowest in these
maps. Complex E lies along the dust spur connecting the two dust lanes
and also has considerable structure.  There are at least two or three
peaks of emission in this complex.  Complex F is an elongated complex
towards the western edge of our field of view; this is the brightest
complex in our field of view.  We also show the boxes used to extract
the spectrum and sum the emission for each complex.  From these, we
calculated the virial and molecular mass for each complex.  The
properties thus derived are listed in Table \ref{tbl-2}.  The method
used for identifying these clouds is discussed in the next section.

\section{Comparing Clouds with Different Cloud Identification Algorithms}

\subsection{The Integrated Intensity Contour Method}\label{just}

We used a specified integrated intensity contour (2 K \kms) to define
a complex.  Is this justified, especially when many of these complexes
clearly have substructure?  From our experience with previous Milky
Way studies, we believe that any cloud complex can always be divided
and sub-divided into smaller and smaller clouds because the molecular
medium is extremely clumpy with a low volume filling factor (see e.g.,
\citealt{dame01}).  The higher the resolution, the more likely it is
that a given structure may be divided into smaller components.  Having
said that, however, a complex has definite boundaries.  Therefore the
integrated intensity contour in a collapsed cube (i.e., collapsed
along one axis) seems to be a reasonable way of identifying a
molecular cloud complex and evaluating its properties.  This has been
applied to GMC studies in the Milky Way \citep[e.g.,]{sanders85}.
Thus we can compare M31 clouds directly to the previous Milky Way
studies.  As a check on the method, we applied the method to a few
Milky Way complexes (Gem OB1, W3CO, Cas A) and to a few complexes in
M33.  
%% SNV I PUT QUOTES FOR SIMULATED AND OBSERVED AND MADE A FEW OTEHR CHANGES IN
%% THE FOLLOWING SENTENCES TO TRY TO CLARIFY
We projected the Milky Way clouds to the distance of M31 and simulated
their interferometric observation using the same $uv$ tracks actually
obtained for the BIMA observations of M31.  The details of these
simulations are in the Appendix. Using this approach each of the three
Milky Way complexes was decomposed into two clouds.  We refer to the
Milky Way complexes projected to the distance of M31 and observed with
the interferometer as ``simulated'' clouds for the remainder of this
paper; the reader should keep in mind that these clouds are not
artificial but are actual Milky Way clouds which are being
``observed'' as if they were in Andromeda and were observed with BIMA.
The properties of the complexes thus defined are listed in Table
\ref{tbl-3}.

\begin{figure}
%\epsscale{}
\plotone{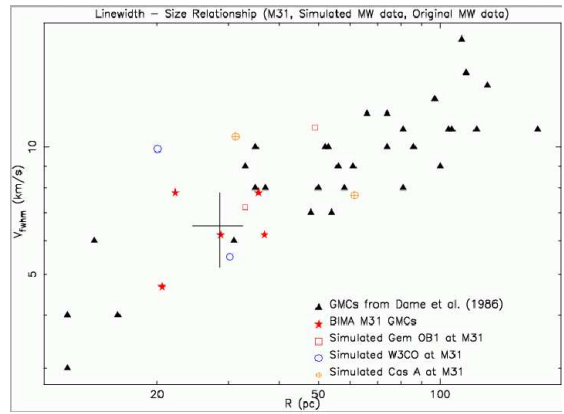}
\caption{The size-linewidth relationship is shown for the M31
  complexes (stars), simulated Milky Way GMCs (squares, circles,
  circles with a plus in the middle) and previously analyzed Milky Way
  complexes (triangles) by \cite{dame86}.  The typical error in
  measurements of the velocity width and radius are shown with the
  cross at the mean value of the M31 clouds in the survey.  All the
  data points fall on the linewidth-size relationship defined by the
  \citet{dame86} Milky Way data.  The data, whether from a single dish
  telescope or an inteferometer, gives equivalent results when the
  observations sample the same spatial scales and have comparable
  sensitivities in different galaxies. \label{fig3}}
\end{figure}
%% SNV THE READER MIGHT BE CONFUSED HERE, SINCE YOU JUST DISCUSSED 'SIMULATED'
%% OBSERVATIONS, YET NOW YOU DISCUSS PROPERTIES DERIVED FROM 
%% ACTUAL MILKY WAY OBSERVATIONS.  SO I THINK WE NEED TO MODIFY TO SAY
%% WE ARE COMPARING 'PROPERTIES' PREVIOUSLY MEASURED, AS i HAVE DONE BELOW
%% When we compare the complexes we observe in M31 to those observed
When we compare the properties of the complexes we observe in M31 to
those observed previously in the Milky Way (e.g.,
\citealt{dame86,sanders85}), we find that the mean properties of the
complexes in the two galaxies are not significantly different.
Whereas \citet{dame86} found complexes which ranging from 24--262 pc
in diameter with velocity widths of 4--18 \kms, the complexes in M31
range from 40--75 pc and 4.7--7.8 \kms.  In Figure \ref{fig3} we place
our M31 complexes, the Milky Way clouds from \citet{dame86}, and a set
of simulated Milky Way clouds on the size-linewidth plot from
\citet{dame86}.  The \citet{dame86} GMCs are particularly useful for
this figure because they have a larger dynamic range in size and
linewidth than the other sets of clouds.  Within the errors, the M31
complexes fall along the same relationship found in the much more
extensive Milky Way survey.  It is also worth noting that the Milky
Way GMCs simulated at the distance of Andromeda and observed with an
interferometer show linewidths and sizes that are also consistent with
the standard relationship.  The main point of this figure is that data
for different galaxies whether observed with a single dish telescope
or an interferometer can be compared equivalently so long as the
observations have comparable spatial resolution (i.e., the
observations sample the same spatial scales in different galaxies) and
sensitivity, and the data are analyzed with the same cloud
identification technique(s).

For identifying clouds in M33, \citet{wilson90} did not use the same
cloud identification method as described above.  In their method, if a
cloud could be resolved by eye in a channel and separated spatially or
in velocity from an adjacent cloud, then the cloud was considered to
be a distinct cloud.  The effect of this technique is similar to the
automated algorithms like Clumpfind\footnote{In fact, when we used
  Clumpfind on the Wilson \& Scoville data, we found every one of
  their clouds, but Clumpfind actually merged 4 pairs of these clouds
  into a single cloud.} and Gaussclump which we discuss in more detail
later in \S \ref{methods}.  Basically, the \citet{wilson90} method
breaks apart complexes whenever possible.  Thus it was difficult to
deduce from their analysis whether the M33 clouds are indeed a
different population than the Milky Way clouds.

We re-analyzed Figure 2 from \citet{wilson90} using the integrated
intensity contour method and found that several of the M33 clouds
merged into larger complexes.  The main reason for that is that the
lowest contour shown in Figure 2a of \citet{wilson90} is 7 K \kms, a
very high emission level. When we use a 2 K \kms contour, clouds MC-6,
MC-7 and MC-8 in field 4C become one complex with a total molecular
mass of 4.3$\times$10$^5$\Msun.  Similarly, in field 4B clouds 11 and
12 (M$_{tot}$=2.4$\times$10$^5$\Msun) and in field 2C clouds 16 and 17
(M$_{tot}$=2.1$\times$10$^5$\Msun) become single complexes.  Also, all
the clouds (except cloud 3 which is at a very different velocity) in
Field 1A merge into one complex.  With better sensitivity, clouds 1,
2, 4 and 5 might be seen as belonging to one complex which would have
a total mass of at least 4.9$\times$10$^5$\Msun.  Although the
difference in the mass of the merged clouds is not significantly
different, the earlier conclusion that no cloud larger than
4$\times$10$^5$\Msun\ exists in M33 appears to be due to a different
choice in cloud identification method and the low sensitivity of the
observations.  Our analysis is intended to show the need for using the
same cloud identification and analysis technique for comparing cloud
properties.

In summary, though we only have a handful of clouds to compare with
the Milky Way, we find that the M31 and M33 GMCs are similar in most
respects to those found in the Milky Way.  The small number of clouds
prevents us from studying the mass function of GMCs in M31. 

\subsection{Automated Clump Finding Algorithms}\label{methods}

Automated algorithms such as Clumpfind \citep{williams94} and
Gaussclump \citep{stutzki90} work on the opposite philosophy from the
single contouring algorithm described above: both Clumpfind and
Gaussclump break apart a complex into as many parts as possible.
Clumpfind contours the cube into discrete intervals specified by the
user.  Then it works its way through the contouring levels identifying
isolated peaks as possible individual clumps.  Clumpfind then assigns
all pixels above the lowest contouring interval to one or the other
clump.  Finally it merges clumps which are closer than the spatial or
velocity resolution.  The advantage of this algorithm is that it has
no preconceptions of the shape of the clouds and hence readily deals
with the complex structure of GMCs.  On the other hand, the contouring
technique tends to clip away too many pixels which can lead to
incorrect calculation of sizes and linewidths.  This may be
particularly problematic for data cubes with low dynamic range.
Gaussclump avoids such problems of noisy or low dynamic range data by
directly fitting clouds in the data cube.  It tries to fit peaks of
emission with an 11-parameter gaussian, also taking into account the
resolution of the observations.  The biggest disadvantage of this
method is that it is restricted to defining clumps as gaussian, which
is seldom the shape of the observed molecular emission.

Neither method is particularly well-suited to analyse GMC properties.
The reason has more to do with molecular clouds than the algorithms
themselves.  Molecular cloud structure is extremely complex, at least
as observed in the Milky Way.  These algorithms tend to divide a
complex into as many subdivisions as allowed by the resolution of the
data.  Hence these methods seldom find structures which are
larger than than one or two resolution elements.  For example, if the
resolution of a survey is 5 pc, these methods will not find a 100 pc
complex because peaks of emission located more than 5 pc apart will be
classified as separate clouds.  The bias thus introduced is not fatal
because cloud/clump properties in equivalent datasets (i.e.  datasets
with similar resolutions and noise characteristics) may still be
compared, but generalizations about a galaxy's cloud population can
not be drawn.  This is especially pertinent to extragalactic astronomy
where higher and higher resolution observations are continually being
achieved with improvements in millimeter wave interferometry.

We advocate that since molecular clouds generally have fairly sharp
edges, the simplest cloud identification algorithms is the integrated
intensity contouring method where one identifies the contour level
associated with the edge of the cloud (e.g.,
\citealt{sanders85,dame86}) in a 3~dimensional cube.  The obvious
shortcoming of this method is that it ignores the structure in the
clouds.  Another obvious disadvantage of the method is that it is
difficult to automate and is quite time consuming because one has to
make three different moment maps along the different axes and decide
what the distinct complexes are.  This method may suffer from blended
emission but this confusion will be limited in extragalactic sources.
\citet{rosolowsky06b} has recently introduced a new method for
identifying and analyzing molecular clouds which may avoid many of the
biases inherent in other methods discussed above; this method may be a
useful tool for future surveys of molecular clouds in nearby spirals
with CARMA and ALMA.

\subsection{M31, M33 and Milky Way Cloud Properties using Gaussclump}\label{auto}

We ran Gaussclump on all CO datasets in all three galaxies.  We find
17 clouds in our BIMA M31 survey, 35 clouds in the M33 survey of
\citet{wilson90}, 7 clouds in the simulated Gem OB1 complex, 5 clouds
in the W3CO complex, and 13 clouds in the Cas A complex.  These
results should be compared to the six complexes found in M31 and the
two complexes found in each of the simulated Milky Way clouds using
the integrated intensity contouring method.The mean values for the
amplitudes, velocity widths and sizes are listed in Table \ref{tbl-4}.
The velocity widths and sizes have not been deconvolved.  Note that
the M33 clouds were not simulated at the distance of M31 and observed
with BIMA.  These data are similar to M31 but many fields have a
slightly coarser resolution (8-9$\arcsec$) than the BIMA data.

\begin{figure}
\epsscale{1.1}
\plotone{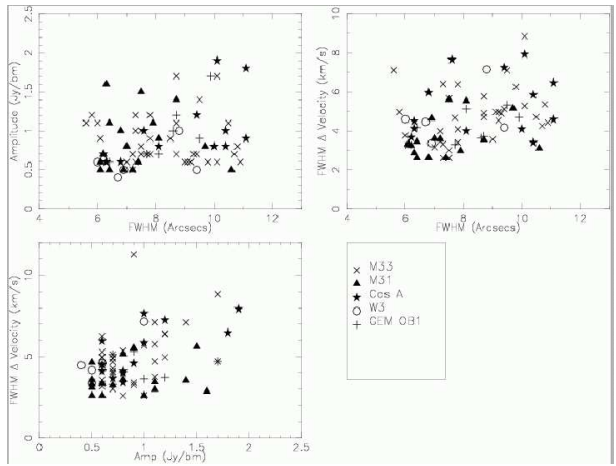}
\caption{The three panels show the three derived properties compared
  with each other for each of the datasets: M31, M33 and the simulated
  Milky Way clouds, Cas A, Gem OB1 and W3CO.  No attempt has been made
  to deconvolve the linewidths or radii because all datasets have the
  same linewidths and approximately the same angular resolution.  The
  sizes are in arcseconds, linewidths in \kms and amplitude in Jy \bm.
  There is no significant difference in the properties of the clouds
  identified by Gaussclumps. The largest M33 clouds are the least
  affected by the coarse ($\sim$ 9$\arcsec$)resolution.  The cluster
  of small M31 clouds with low linewidths are those in Complex C.
  These are very weak clouds and their properties are not
  well-defined.
  \label{fig5}}
\end{figure}

The derived properties are also shown graphically in Figure
\ref{fig5}.  The plots show no significant differences in the
amplitudes, sizes or velocity widths for the three galaxies.  We plot
each of the derived properties against another derived property to
test whether there is any dependence of one on the other, and find
none.  

Our conclusion from this experiment is similar to that derived from
the analysis of the integrated intensity contouring method.  The
molecular clouds in these datasets, when compared in a consistent
manner, show no significant differences.

\section{Conclusions}\label{conc}
In the northeastern spiral arm of M31, we have identified six
distinct, large complexes of molecular gas.  All of these complexes
lie along the spiral arm dust lanes, or in one case (Complex E), along
a dust spur in between the dust lanes.  These complexes have a mean
diameter of 57$\pm$13 pc, a mean velocity width of 6.5$\pm$1.2 \kms,
and a mean molecular mass of 3.0$\pm$1.6 $\times$ 10$^5$ \Msun and are
indistinguishable from those found in the Milky Way e.g.,
\citet{dame86}.

Meaningful comparison of GMCs in different galaxies requires consistent
analysis of data taken with different instruments and different cloud
identification techniques.  This paper represents an attempt at
eliminating such differences and comparing GMCs in M31, M33 and the
Milky Way in as consistent a manner as possible.
We have simulated three Milky Way complexes at the distance of M31 and
observed them with the BIMA array.  The simulations show that
interferometers are excellent at recovering molecular complexes in the
Local Group galaxies.  We compared the simulated Milky Way complexes
to the M31 data using an integrated intensity contour method and
found that the complexes thus identified and analyzed fell on the same
linewidth size relationship as found by \citep{dame86}.  The M33
clouds if analyzed in this manner also yield larger clouds than
previously stated.  Larger interferometric surveys of these galaxies are
necessary to compare the GMC distribution and mass function to that in
the Milky Way (e.g., \citealt{engargiola03, rosolowsky07a}).  

Finally, we compared two automated algorithms, Clumpfind and Gaussclump,
and found that Gaussclump was better at recovering cloud properties
than Clumpfind for low dynamic range data.  Our experiments revealed
that such algorithms are resolution dependent and the inherent clumpy
nature of molecular emission prevents these methods from identifying
clouds larger than one or two resolution elements.  Hence we caution
against using these methods to characterize GMC populations
in galaxies.  Still, these may be used to quickly compare properties
of clouds in data with similar noise and resolution characteristics.
Using Gaussclump, we find that the cloud properties in M31, M33 and
simulated Milky Way data are indistinguishable. 

\acknowledgments We are grateful for the insightful and helpful
comments by the referee Elly Berkhuijsen which greatly improved the
manuscript.  We thank Leo Blitz, Greg Engargiola, Tamara Helfer,
Andrew Harris, Lee Mundy, Marc Pound, Erik Rosolowsky, Peter Teuben,
Stephen White and Friedrich Wyrowski for helpful discussions.  We
acknowledge the use of NASA's {\it SkyView} facility
(http://skyview.gsfc.nasa.gov) located at NASA Goddard Space Flight
Center.  We also acknowledge support for the BIMA array from the
National Science Foundation under grant AST-9981289 to U. Maryland.

%% In this section, we use  the \subsection command to set off
%% a subsection.  \footnote is used to insert a footnote to the text.


%% This section also includes several of the displayed math environments
%% mentioned in the Author Guide.
\centerline{\bf{Appendix}}
\appendix
%% SNV I THINK TITLE NEEDS WORK
%% \section{Simulating Milky Way at M31 for BIMA observations}\label{simul}
\section{Simulation of BIMA Observation of Milky Way GMCs at the Distance of M31}

The first step is to simulate the appearance of the Milky Way clouds
if they were located at the distance of Andromeda.  In other words, we
need to scale the Milky Way maps to the distance of M31, which we did
by scaling the map header values for the angular size of the map
pixels by the ratio of the distance to each Milky Way complex to that
of M31.  The next step is to simulate observation of the
distance-scaled Milky Way data with the BIMA interferometer.  To get a
precise match in the point-spread function, we modified the header
values for the RA and Dec at the field center so as to match the
position of the BIMA M31 observation; this enables the $uv$ tracks for
the simulated observations to match the actual observations.  The
original temperature scale which was in the units of T${_A^*}$ was
converted to Jy \bm and the maps were gridded to a 2$\arcsec$ pixel
scale. A separate model was then created for each mosaicked pointing;
each model had a primary beam taper applied to it.  All seven of these
models were then observed with the {\it uv \rm} tracks used for M31.
Since the original Milky Way data were approximately half as noisy as
the BIMA data, we simulated additive noise in the {\it uv \rm} plane
and added that to our final maps.  The final simulated Milky Way maps
are therefore reflective of the effects of spatial filtering by the
interferometer, and are similar in almost every way to the M31 and M33
data.

%% SNV TITLE ALSO NEEDS WORK (I HOPE BIMA DOESN''T AFFECT CARINA!)
%%\section {Effects of the interferometer  on the Carina, Gem OB1, W3CO,
%%  and Cas A cloud complexes}\label{carina}
\section{Effect of Interometric Filtering on Simulations of Milky Way
Observations}

\begin{figure}
\plotone{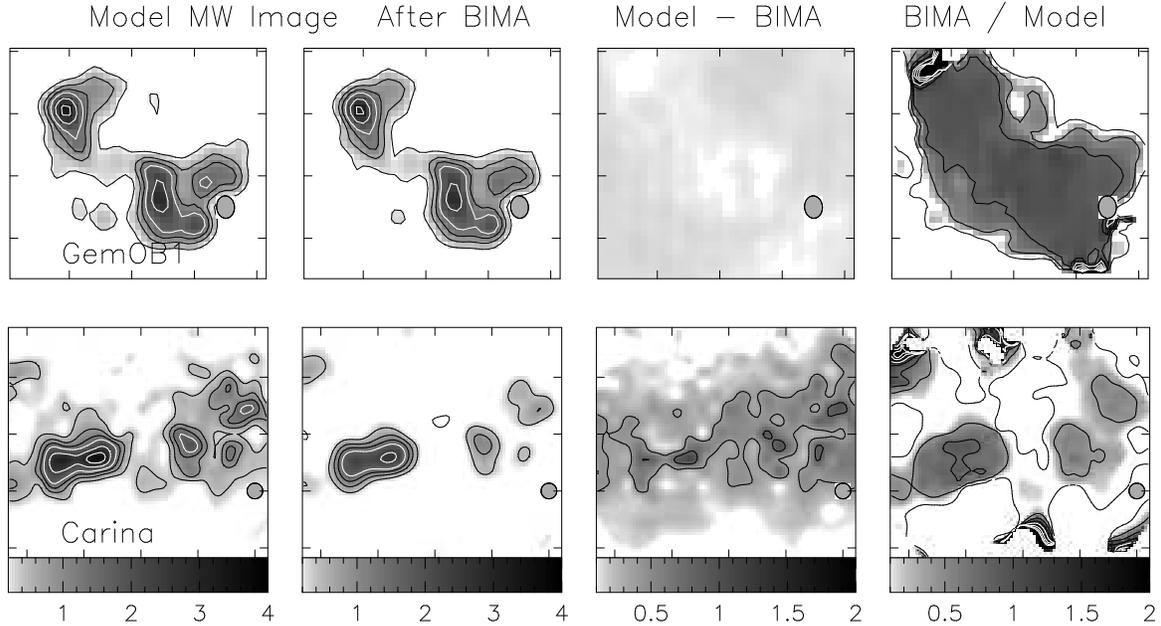}
% \plotfiddle{f6.eps}{5in}{-90}{}{}{}{}
\caption{The top row of this figure shows a channel from the simulation of the Gem OB1 complex and the bottom row 
  shows a channel from the simulation of the Carina complex.  The first column is
  the Milky Way cloud at the distance of M31.  The second column is the
  cloud simulated with interferometric observations with BIMA.  The grey scale wedge below the rows is an indication of the flux in Jy \bm for the first three columns; it is a ratio for the final column.   In column 2, one can immediately notice that the Carina cloud is significantly dimmer than its input model whereas the Gem OB1 cloud appears unchanged.  The third column is the model cloud minus the simulated image, and the fourth column is the simulated image divided by the model image.  Column 3 is in Jy \bm.  Notice that almost no flux is lost in the
  Gem OB1 simulation whereas Carina arm cloud is weaker by as much as
  30-50 \%. \label{fig6}}
\end{figure}
We initially began this comparative study project with a section of
the Carina arm from the 1.2m survey because, at a heliocentric
distance of 13.6 kpc, the Milky Way survey's linear resolution matched
the BIMA observations of M31.  However when we simulated the Carina
clouds, the simulated maps showed significant differences as shown in
Figure \ref{fig6}.  The bottom  row of this figure shows a single
velocity channel from the Carina simulations.  The first column is
the model Carina complex at the distance of M31 before BIMA
observations.  The second column is the model after BIMA observations,
i.e. the simulated map.  Both of these columns are on the same greyscale
and have the same contour levels.  Column 3 is the input model minus
the simulated map, and column 4 is the simulated map divided by the
input model. One immediately sees the reduction in the flux in these
maps.  The clouds are dimmer by a factor of 30-50\%.  The flux that is
resolved out is not a uniform background but rather structure on all
scales. We further tested these effects by changing the orientation of
the Carina model clouds by shifting them in different directions and
rotating them at various angles.  In every case, we saw the same
result: the flux was reduced by as much as 50\%.

When we repeated the experiments for the Cas A, W3CO and Gem OB1
clouds from the same Milky Way Survey, we found a completely different
result.  The interferometer recovers almost all of the flux for these
simulations.  An example of this is shown in the top row, which shows
a single velocity channel from the Gem OB1 simulations.  The
difference between the Carina clouds and these complexes is that these
complexes are 4-6 times closer than Carina and were observed with a
linear resolution of $\sim$5 pc compared to a resolution of $\sim$30
pc.  The 1.2m maps of the Carina clouds therefore have smoother
emission on a larger scale than the Cas A, W3CO and Gem OB1 clouds;
this smooth emission is then subsequently resolved out by the interferometer.  This
filtering of flux from large, smooth structures by interferometers is
expected and observed (e.g., \citealt{beck96, helfer02}).  Presumably,
the Carina arm at higher resolution would look like the other
Milky Way complexes we studied (Gem OB1, W3CO and Cas A).  Since the
Carina arm model itself is incorrect, we cannot use the simulated
Carina clouds in our comparative analysis.  The recovery of the Gem
OB1, W3CO and Cas A clouds is reassuring because it means that
interferometers are excellent at recovering the molecular emission in
GMCs; this is due to the inherently clumpy nature of GMCs.

% \begin{figure}
% \plotone{f4.eps}
% \caption{Comparison of Gaussclump and Clumpfind.
 % The y-axis is the value derived by the algorithms in the model data
 % before the interferometer observations and the x-axis is the same
 % value derived after the filtering through BIMA.  The top row shows
 % the results from Gaussclump and the bottom row shows results of
 % Clumpfind in the Carina arm data. Notice the almost systematic
 % reduction in size and velocity widths of most clumps in Clumpfind.
 % \label{fig4}}
% \end{figure}

\begin{deluxetable}{ll}
\tabletypesize{\scriptsize}
\tablecaption{Observing Parameters for M31 \label{tbl-1}}
\tablewidth{0pt}
\tablehead{\colhead{} & \colhead{}}
\startdata
~~$\alpha$(J2000)\tablenotemark{a} & 0$^h$45$^m$07$^s$.57 \\
~~$\delta$(J2000)\tablenotemark{a} & 41$^\circ$35$^\prime$46$\farcs$68 \\
V$_{LSR}$( \kms) & -135\\
Total bandwidth (\kms) & 142.85 \\
Velocity resolution (\kms) & 0.254\tablenotemark{b} \\
Projected baselines (k$\lambda$ & 2.2-30.2 \\
Single-sideband T$_{SYS}$ (Kelvin)& 275 - 1000  \\
Calibration quasar &  0136+478 \\
Absolute flux calibrator & Uranus \\
Typical uncertainty in flux calibration & 15\% \\
Beam & 7$\farcs$11 $\times$ 5$\farcs$71 $\times$ 4.7$^\circ$ \\
Noise in 2.6 \kms channel (Jy \bm) & 0.15 \\
\enddata
\tablenotetext{a}{Central pointing}
\tablenotetext{b}{Correlator data taken at 0.254 \kms\ resolution.  Data were later smoothed to 2.6 \kms\ resolution for imaging.}
\end{deluxetable}

\begin{deluxetable}{llll}
\tabletypesize{\scriptsize}
\tablecaption{Datasets \& Instruments \label{tbl-1a}}
\tablewidth{0pt}
\tablehead{\colhead{Galaxy} & \colhead{Observatory}& \colhead{Angular/Linear Resolution}& \colhead{Assumed distance}}
\startdata
~~M31 &  BIMA &  7$\farcs$11 $\times$ 5$\farcs$71 / 19-24 pc & 690 kpc \\
~~M33 & OVRO & 5--7$\arcsec$ /  19--26 pc & 790 kpc \\
~~GEM OB1 (Milky Way) & CfA 1.2m & 8.8$\arcmin$ / 5.1 pc& 2 kpc \\
~~Cas A (Milky Way) & CfA 1.2m & 8.8$\arcmin$ 5.1 pc / & 2 kpc \\
~~W3 (Milky Way) & CfA 1.2m & 8.8$\arcmin$ / 6.4 pc & 2.5 kpc \\
~~Carina\tablenotemark{a} (Milky Way) & CfA 1.2m & 8.8$\arcmin$ / 34.8 pc & 13.6 kpc \\
\enddata
\tablenotetext{a}{Carina data not used in comparative study}
\end{deluxetable}

\begin{deluxetable}{cccccccccc}
\tablecolumns{10}
\tabletypesize{\scriptsize}
\tablecaption{Properties of M31 GMCs\label{tbl-2}}
\tablewidth{0pt}
\tablehead{\colhead{Cloud} & \colhead{RA} & \colhead{DEC} &
  \colhead{Vpeak} & \colhead{V$_{fwhm}$\tablenotemark{a, b}} & \colhead{T$_B^*$\tablenotemark{c}}&
  \colhead{S$_{CO}$\tablenotemark{a}} & \colhead{Diameter\tablenotemark{a, d}} & \colhead{M$_{VIR}$\tablenotemark{a, e}}  &  \colhead{M$_{mol}$\tablenotemark{a, f}}  \\
\colhead{} & \colhead{(J2000)} & \colhead{(J2000)} & \colhead{\kms} &
  \colhead{\kms} & \colhead{K} & \colhead{Jy \kms} & \colhead{pc} &
  \colhead{$\times$10$^5$\Msun} & \colhead{$\times$10$^5$\Msun}}  
\startdata
A & 0:45:12.42 & 41:37:07.23 & -126 & 7.8 & 3.32 & 40.8 & 44.5 &
2.68 & 3.13 \\
B & 0:45:12.73 & 41:35:24.49 & -139 & 7.8 & 3.61 & 67.6 & 71.2 &
4.29 &  5.18 \\
C & 0:45:08.69 & 41:27:49.58 & -134 & --  & 1.57  & --  & --  &
--  & -- \\
D & 0:45:06.04 & 41:35:23.25 & -134 & 4.7 & 2.02 & 11.1 & 41.1 &
0.89 &  0.85 \\
E & 0:45:04.97 & 41:36:11.69 & -126 & 6.2 & 3.34 & 55.4 & 73.6 &
2.80 &  4.25 \\
F & 0:45:01.96 & 41:36:38.99 & -131 & 6.2 & 4.24 & 23.6 & 57.4 &
2.2 & 1.80 \\
\enddata
\tablenotetext{a}{Typical errors are:  D,$\pm$ 4pc, V$_{fwhm}$,$\pm$ 1.3 \kms, S$_{CO}$,$\pm$20\%, M$_{VIR}$,$\pm$30\%, M$_{mol}$,$\pm$30\%}
\tablenotetext{b}{Full width at half maximum measured from the spectrum}
\tablenotetext{c}{Brightest pixel in channel maps}
\tablenotetext{d}{Average of deconvolved major and minor axis gaussian
  fit to integrated intensity map of complex}
\tablenotetext{e}{M$_{VIR}$= 99 $\times$ V$_{fwhm}^2$ $\times$ Diameter}
\tablenotetext{f}{M$_{mol}$ = 1.61 $\times$ 10$^4 \times$ d$^2 \times$ S$_{CO}$ This assumes a Galactic value for the conversion factor between the CO flux and the hydrogen column density of 3$\times$ 10$^{20}$ cm$^{-2}$ (K \kms)$^{-1}$ 
and includes a factor of 1.36 to account for helium and other heavy elements.}\end{deluxetable}

\begin{deluxetable}{ccccccc}
\tablecolumns{7}
\tabletypesize{\scriptsize}
\tablecaption{ Properties of Milky Way GMCs Simulated at M31 \label{tbl-3}}
\tablewidth{0pt}
\tablehead{\colhead{Cloud} & \colhead{V$_{fwhm}$\tablenotemark{a, b}} & \colhead{T$_B^*$\tablenotemark{c}}&
  \colhead{S$_{CO}$\tablenotemark{a}} & \colhead{Diameter\tablenotemark{a, d}} & \colhead{M$_{VIR}$\tablenotemark{a, e}}  &  \colhead{M$_{mol}$\tablenotemark{a, f}}  \\
\colhead{} &
  \colhead{\kms} & \colhead{K} & \colhead{Jy \kms} & \colhead{pc} &
  \colhead{$\times$10$^5$\Msun} & \colhead{$\times$10$^5$\Msun}}  
\startdata
Gem1 & 7.2 &  4.0 & 70.87 & 66.0 & 3.4 & 5.4 \\
Gem2 & 11.1 & 2.9 & 105.4 & 98.2 & 12.0 & 8.0 \\ 
W1   & 9.9 & 2.4 & 42.62 & 40.1 & 3.9 & 3.3 \\
W2   & 5.5 & 1.3 & 22.21 & 61.5 & 1.8 & 1.7 \\
CasA1 & 7.7 & 4.5 & 260.9 & 123.0 & 7.2 & 20 \\
CasA2 & 10.6 & 2.9 & 81.8 & 62.5 & 6.9 & 6.3 \\
\enddata
\tablenotetext{a}{Typical errors are:  D,$\pm$ 4pc, V$_{fwhm}$,$\pm$ 1.3 \kms, S$_{CO}$,$\pm$20\%, M$_{VIR}$,$\pm$30\%, M$_{mol}$,$\pm$30\%}
\tablenotetext{b}{Full width at half maximum measured from the spectrum}
\tablenotetext{c}{Brightest pixel in channel maps}
\tablenotetext{d}{Average of deconvolved major and minor axis gaussian
  fit to integrated intensity map of complex}
\tablenotetext{e}{M$_{VIR}$= 99 $\times$ V$_{fwhm}^2$ $\times$ Diameter}
\tablenotetext{f}{M$_{mol}$ = 1.61 $\times$ 10$^4 \times$ d$^2 \times$ S$_{CO}$ }
\end{deluxetable}

\begin{deluxetable}{ccccc}
\tablecolumns{5}
\tabletypesize{\scriptsize}
\tablecaption{Mean of Gaussclump Derived Properties\label{tbl-4}}
\tablewidth{0pt}
\tablehead{\colhead{Galaxy} & \colhead{\# of clouds} & \colhead{Amplitude} &
  \colhead{V$_{fwhm}$} & \colhead{Size} \\
\colhead{} & \colhead{} & \colhead{Jy beam$^{-1}$} & \colhead{\kms} &
  \colhead{Arcseconds}}

\startdata
M31 & 17 & 0.9$\pm$0.4 & 3.6$\pm$1.0 & 7.4$\pm$1.3 \\
M33 & 35 & 0.9$\pm$0.3 & 4.9$\pm$1.7 & 8.4$\pm$1.4 \\
Gem OB1\tablenotemark{a} & 7 & 1.0$\pm$0.4 & 4.3$\pm$0.7 & 8.4$\pm$1.1 \\
W3CO\tablenotemark{a}  & 5 & 0.6$\pm$0.2 & 4.7$\pm$1.3 & 7.6$\pm$1.3 \\
Cas A\tablenotemark{a}  & 13 & 1.0$\pm$0.4 & 5.3$\pm$1.5 & 8.8$\pm$1.8 \\
\enddata
\tablenotetext{a}{These values refer to the simulated Milky Way clouds at the distance of M31.} 
\end{deluxetable}

\end{document}